\documentstyle[psfig,nicV]{article}
\newcommand{\msun}{M$_\odot$ }

\newcommand{\nampgmg}{$^{21}$Na(p,$\gamma$)$^{22}$Mg }
\newcommand{\natomg}{$^{23}$Na(p,$\gamma$)$^{24}$Mg }
\newcommand{\altosi}{$^{25}$Al(p,$\gamma$)$^{26}$Si }
\newcommand{\algtosi}{$^{26}$Al$^{g}$(p,$\gamma$)$^{27}$Si }
\newcommand{\almtosi}{$^{26}$Al$^{m}$(p,$\gamma$)$^{27}$Si }
\newcommand{\sitop}{$^{26}$Si(p,$\gamma$)$^{27}$P }
\newcommand{\altomg}{$^{27}$Al(p,$\alpha$)$^{24}$Mg }
\newcommand{\mgtoal}{$^{26}$Mg(p,$\gamma$)$^{27}$Al }
\newcommand{\napgmg}{$^{22}$Na(p,$\gamma$)$^{23}$Mg }
\newcommand{\mgpgal}{$^{23}$Mg(p,$\gamma$)$^{24}$Al }
\newcommand{\alsmall}{$^{23}$Al(p,$\gamma$)$^{24}$Si }

\newcommand{\na}{$^{22}$Na }

\newcommand{\mggg}{$^{26}$Mg }

\newcommand{\physrep}{Phys. Rep.\,\,}
\newcommand{\physrevl}{Phys. Rev. Let.\,\,}
\newcommand{\physrev}{Phys. Rev.\,\,}

\newcommand{\nucphys}{Nuc. Phys.\,\,}
\newcommand{\ADNDT}{ADNDT\,\,}

\begin{document}
%
%
\heading{%
Nuclear uncertainties in the NeNa-MgAl cycles and \\
synthesis of ${}^{22}$Na and ${}^{26}$Al in classical novae\\
}
\par\medskip\noindent
%
\author{%
Jordi Jos\'e$^{1,2}$, Alain Coc$^{3}$, Margarita Hernanz$^{2}$
}
\address{
Departament de F\'{\i}sica i Enginyeria Nuclear (UPC), Av.
 V\'{\i}ctor Balaguer s/n, E-08800 Vilanova i la Geltr\'u (Barcelona), SPAIN
}
\address{
 Institut d'Estudis Espacials de Catalunya (IEEC/CSIC/UPC), 
 Edifici Nexus-201, Gran Capit\`a 2-4, E-08034 Barcelona, SPAIN
}
\address{
 Centre de Spectrom\'etrie Nucl\'eaire et de Spectrom\'etrie de
 Masse, IN2P3-CNRS, B\^at.104, F-91405 Orsay Campus, FRANCE
}
\begin{abstract}
   Classical novae eject significant amounts of
 matter into the interstellar medium, as a result of thermonuclear
 runaways. Nucleosynthesis associated with nova outbursts includes
 products from explosive H-burning, such as $^{17}$O, $^{15}$N and $^{13}$C, 
 and also radioactive species like $^7$Be, $^{22}$Na and $^{26}$Al. 
 In this paper we report on new hydrodynamic calculations of nova
 outbursts, from the onset of accretion to mass ejection. We stress the
 role played by the nuclear uncertainties associated with key reactions of 
 the NeNa-MgAl cycles on the synthesis of $^{22}$Na and $^{26}$Al. 
\end{abstract}
\section{Introduction}
The thermonuclear runaway model has been successful in reproducing the
main observational properties of classsical nova outbursts. 
According to this scenario, novae are produced by thermonuclear
runaways located in the white dwarf component of
a close binary system. The large, main sequence companion overfills 
its Roche lobe, providing matter outflows through the inner Lagrangian point 
that lead to the formation of an 
accretion disk around the white dwarf. A fraction of this H-rich matter 
lost by the companion ultimately ends up on top of the white dwarf, where it 
is gradually compressed as accretion goes on. The piling up of matter heats 
the envelope up to the point when ignition conditions to drive a
TNR are reached.

An extended set of hydrodynamic computations of classical nova outbursts has
been performed during the last 25 years (see \cite{Sta98}, \cite{KP97}, 
\cite{JH98}, and references therein), for a wide range of white dwarf masses 
and initial luminosities, which have been able to
identify two types of nova outbursts, those occurring in CO and in
ONe white dwarfs. The latter ones have provided a framework for the origin 
of the high concentrations of Ne and
more massive isotopes found in the spectra of some well-observed nova systems
\cite{LT94}. Among the isotopes synthesized in 
these so-called ONe novae (mainly $^{13}$C, ${}^{17}$O and ${}^{15}$N), two
radioactive species have raised a particular astrophysical interest: 
${}^{22}$Na and ${}^{26}$Al. 
In this paper we investigate the role played by the nuclear uncertainties 
associated with key reactions of the NeNa-MgAl cycles on the synthesis of
${}^{22}$Na and ${}^{26}$Al during classical nova outbursts.
\section{Synthesis of $^{22}$Na and $^{26}$Al in nova outbursts}
The role of \na as a diagnosis of a nova outburst was first 
suggested by \cite{CH74}. In its decay to a short-lived excited 
state of ${}^{22}$Ne ($\tau \sim 3.75$ yr), \na emits a 
$\gamma$-ray photon of $1.275$ MeV. Through this mechanism, nearby ONe novae 
within a few kiloparsecs from the Sun may provide detectable $\gamma$-ray 
fluxes. Several experimental verifications of this $\gamma$-ray emission at
1.275 MeV from nearby novae have been attempted in the last twenty years, 
from which upper limits on the ejected ${}^{22}$Na have been derived.  
 In particular, the 
observations performed with the COMPTEL experiment on-board CGRO of
five recent neon-type novae \cite{Iyu95}, as well as observations
of {\it standard} novae, have led to an upper limit of
$3.7 \times 10^{-8}$ \msun for the ${}^{22}$Na mass ejected by any nova in 
the Galactic disk. This limit has posed some constraints on 
pre-existing theoretical models of classical nova explosions. 

${}^{26}$Al is another unstable nucleus, with a lifetime of $\tau = 1.04 \times 
10^6$ years, that decays from ground state to the first excited level of \mggg,
which in turn de-excites to its ground state by emitting a gamma-ray photon 
of 1.809 MeV.  This characteristic gamma-ray signature, first detected 
in the Galatic Center by the HEAO-3 satellite, has been confirmed by other 
space missions.
The most recent measurements made with COMPTEL have provided a map of the 
1.809 MeV emission in the Galaxy (\cite{Die95}, \cite{PD96}). 
The inferred $1-3$ \msun of Galactic ${}^{26}$Al are, according to the
observed distribution, mainly attributed to young progenitors, such as 
massive AGB stars, SN II and Wolf-Rayet stars. However, a 
potential  contribution from novae or low-mass AGB stars cannot be ruled out.  

First estimates of the ${}^{22}$Na and ${}^{26}$Al production in novae  
were performed by different groups, using simplified one-zone 
models with representative temperature and density profiles
 (see \cite{HT82}, \cite{Wie86}, \cite{WT90}, \cite{NSS91}), on the basis of 
ONeMg white dwarf stars. This scenario was revisited by \cite{Pol95} using 
hydrodynamic computations: the obtained ${}^{22}$Na yields range between
$5 \times 10^{-5}$ and $5 \times 10^{-3}$, by mass.
Assuming that the whole envelope ($\sim 10^{-4}-10^{-5}$ \msun) is ejected 
during the outburst, 
they derive peak fluxes of the ${}^{22}$Na 1.275 MeV line in the range 
$(2-30)\times 10^{-5}$ photons s$^{-1}$ cm$^{-2}$ (corresponding to ONeMg 
white dwarf masses between $1.00-1.35$ \msun, respectively) for  
classical novae exploding at 1 kpc.
Their main conclusion, based on the reported $3 \sigma$ line sensitivity of
OSSE and COMPTEL, is that
nearby classical novae involving ONeMg white dwarfs of $M_{wd} \geq 1.25$ \msun
should produce detectable ${}^{22}$Na $\gamma$-rays for CGRO, a prediction
not confirmed so far \cite{Iyu95}.
 Their results showed also a significant production of ${}^{26}$Al 
(i.e., $(19.6-7.5) \times 10^{-3}$, by mass, for the same white dwarf mass
 range), which could
account for a major fraction of the Galactic ${}^{26}$Al. 

Recent hydrodynamic computations (\cite{JHC97}, \cite{JH97}, \cite{JH98})
using updated nuclear reaction rates, have led to a significant reduction
of both ${}^{26}$Al and ${}^{22}$Na ejected during nova outbursts.
In particular, a mean mass fraction of $1 \times 10^{-4}$ of ${}^{22}$Na is 
found in the 1.25 \msun ONe Model 
(with $M_{ejec}$(${}^{22}$Na)=$ 1.3 \times 10^{-9} $ \msun),
whereas a maximum value of $6 \times 10^{-4}$ results from the 1.35 \msun 
ONe Model (with $M_{ejec}$(${}^{22}$Na)=$ 2.6 \times 10^{-9}$ \msun). 
The corresponding peak fluxes in the 1.275 MeV ${}^{22}$Na line, below 
$10^{-5}$ counts s$^{-1}$ cm$^{-2}$ for novae at 1 kpc, turn out to be too low 
to be detected with
OSSE or COMPTEL but represent potential targets for the nearby future INTEGRAL
mission \cite{Gom98}. 
Concerning ${}^{26}$Al, yields ranging from $2 \times 10^{-3}$ to 
$2 \times 10^{-4}$ by mass have been
obtained in a series of ONe nova models with masses between $1.15-1.35$ \msun. 
Contribution of novae to the Galactic ${}^{26}$Al is 
limited to $0.1-0.4$ \msun  \cite{JHC97}.

Other hydrodynamic computations performed by \cite{Sta97} and \cite{Sta98},
 using also updated nuclear reaction rates and opacities, have 
modified previous estimates (i.e., \cite{Pol95}). 
 The expected abundance of ${}^{22}$Na in the ejecta has risen up to 
 $(2-3) \times 10^{-3}$, by mass, when 1.25 \msun ONeMg white dwarfs are 
adopted, high enough to be detected by CGRO, provided that all the accreted
envelope ($\sim 3 \times 10^{-5}$ \msun) is ejected. On the other hand, 
the improved input physics translates into a factor of 10 reduction  
on the synthesis of ${}^{26}$Al, in better agreement with the analysis
of the 1.809 MeV emission map provided by COMPTEL, and also
with the results previously reported by \cite{JHC97}.

Since the synthesis of both ${}^{22}$Na and ${}^{26}$Al is very dependent on
the adopted nuclear reaction rates, we have investigated the effect of
the nuclear uncertainties accompanying reaction rates within the NeNa-MgAl
cycles on the synthesis of ${}^{22}$Na and ${}^{26}$Al, using an updated
version of the spherically symmetric, hydrodynamic SHIVA code \cite{JH98}.
In a first step, we have identified the most relevant reactions involved
in the synthesis of such isotopes. Main results are summarized in Table 1.
We stress the crucial role played by \nampgmg: a reduction by a factor
100 of the rate given by \cite{CF88} translates into an increase in the 
mean abundance of ${}^{22}$Na in the ejecta by a factor between 1.2 and 3. 
An increase by a factor of 3 is also obtained when the rate given by 
\cite{Ste96},
instead of that from \cite{CF88}, is adopted for \napgmg. Worth noticing
is also the increase in the final amount of ${}^{26}$Al by a factor 2
when the term corresponding to the 0.188 MeV resonance, as measured by
\cite{Vog89}, is reduced by just a factor 1/3.
\begin{center}
\begin{tabular}{l c c c c}
\multicolumn{5}{l}{{\bf Table 1.} Test Models} \\
\hline
\multicolumn{1}{c}{Nuclear reaction}& \multicolumn{1}{c}{Old rate}&
\multicolumn{1}{c}{Test rate} &
\multicolumn{1}{c}{$\rm \frac{X(^{22}Na)_{test}}{X(^{22}Na)_{old}}$}&
\multicolumn{1}{c}{$\rm \frac{X(^{26}Al)_{test}}{X(^{26}Al)_{old}}$}\\
\hline
\multicolumn{5}{c}{1.15 $M_\odot$ ONe} \\
\hline
\nampgmg & CF88  & CF88/100   &2.3 & 1.1 \\
\altosi  & Wie86 & Coc95, case A& 1& 1.2 \\
         & Wie86 & Coc95, case C& 1& 0.5 \\
\natomg  & CF88  & CF88+Gor89 &1.2 & 1.3 \\
\algtosi & Vog89 & Vog89      & 1  & 1.9 \\
  &    & 1/3$\times$res(0.188 MeV) &    &     \\
\almtosi & CF88  & CF88$\times$100   & 1  &  1  \\
\sitop   & Wie86 & Her95      & 1  &  1  \\
\altomg  & Cha88 & Cha88+Tim88& 1  &  1  \\
\mgtoal  & Ili90 & Ili90+Cha90& 1  &  1  \\
\hline
\multicolumn{5}{c}{1.25 $M_\odot$ ONe} \\
\hline
\nampgmg & CF88  & CF88/100   & 3  & 1.1 \\
\napgmg  & CF88  & Ste96      & 3  & 1.2 \\
\mgpgal  & Wie86 & Kub95      &1.1 & 1.1 \\
\hline
\multicolumn{5}{c}{1.35 $M_\odot$ ONe} \\
\hline
\nampgmg & CF88  & CF88/100   &1.2 &  1  \\
\alsmall & Wor94 & Sch97      & 1  &  1  \\
\hline
\end{tabular}
\end{center}
In a second step, a series
of hydrodynamic nova models has been computed assuming 
upper, recommended and lower estimates of the reaction rates, 
from which limits on the production of both ${}^{22}$Na and ${}^{26}$Al 
are derived (see \cite{JCH98} for details). Main results are summarized in 
Table 2.
Large differences in the $^{22}$Na and $^{26}$Al yields have been 
obtained: in particular, a factor $\sim 4$ in the ${}^{22}$Na production
for the 1.15 \msun ONe Model, and a factor $\sim 7$ in the case of 
${}^{26}$Al production (1.35 \msun ONe Model)  when either 
the upper or the lower estimates are adopted. 
These differences are too large to accurately predict the
contribution of classical novae to the Galactic ${}^{26}$Al, or to 
determine the maximum distance to a classical novae for a potential
detection of the 1.275 MeV ${}^{22}$Na $\gamma$-ray line. 
\begin{center}
\begin{tabular}{l c c c c c}
\multicolumn{6}{l}{{\bf Table 2.} Ranges of $^{22}$Na and $^{26}$Al 
                                  production} \\
\hline
\multicolumn{1}{c}{Model}& \multicolumn{1}{c}{Network}&
\multicolumn{1}{c}{X($^{22}$Na)} &\multicolumn{1}{c}{X($^{26}$Al)}& 
\multicolumn{1}{c}{$\rm \frac{X(^{22}Na)_{upper}}{X(^{22}Na)_{lower}}$}&
\multicolumn{1}{c}{$\rm \frac{X(^{26}Al)_{upper}}{X(^{26}Al)_{lower}}$}\\
\hline
1.15 $M_\odot$ ONe & Lower  & 1.2E-4 & 3.3E-4 & 3.8 & 3.6 \\
                 &   Upper & 4.6E-4 & 1.2E-3 &     &     \\
\hline
1.25 $M_\odot$ ONe & Lower  & 1.9E-4 & 2.0E-4 & 3.2 & 4.4 \\
                 & Upper & 6.0E-4 & 8.7E-3 &     &     \\
\hline
1.35 $M_\odot$ ONe & Lower  & 7.7E-4 & 1.5E-4 & 1.8 & 7.3 \\
                 &   Upper & 1.4E-3 & 1.1E-3 &     &     \\
\hline
\end{tabular}
\end{center}
Therefore, we stress the need of new nuclear physics experiments to 
reduce the uncertainties associated with some key reactions of the
NeNa-MgAl cycles, in particular \nampgmg,  \napgmg  and \altosi. 
A verification of the yet unpublished values corresponding to the
0.188 MeV resonance of \algtosi, measured by \cite{Vog89}, is also
recommended.

\acknowledgements{This research has been partially supported by the
 DGICYT (PB97-0983-C03-02; PB97-0983-C03-03), by the CICYT-P.N.I.E., 
 by the PICS 319, and by a CIRIT grant. }

\vfill
\end{document}